\documentclass[aps,prd,preprint,preprintnumbers,nofootinbib,superscriptaddress]{revtex4}
\usepackage{ae}
\usepackage{bm} 
\usepackage{amssymb}
\usepackage{amsmath}
\usepackage{amsfonts}
\usepackage{slashed}
\usepackage{ulem}
\usepackage{bbold}

\newcommand{\Eqref}[1]{Eq.~\eqref{#1}}

\allowdisplaybreaks

\begin{document}

\title{Derivative corrections to the Heisenberg-Euler effective action}

\author{Felix Karbstein}\email{felix.karbstein@uni-jena.de}
\affiliation{Helmholtz-Institut Jena, Fr\"obelstieg 3, 07743 Jena, Germany}
\affiliation{GSI Helmholtzzentrum f\"ur Schwerionenforschung, Planckstra\ss e 1, 64291 Darmstadt}
\affiliation{Theoretisch-Physikalisches Institut, Abbe Center of Photonics, \\ Friedrich-Schiller-Universit\"at Jena, Max-Wien-Platz 1, 07743 Jena, Germany}

\date{\today}

\begin{abstract}
We show that the leading derivative corrections to the Heisenberg-Euler effective action can be determined efficiently from the vacuum polarization tensor evaluated in a homogeneous constant background field.
After deriving the explicit parameter-integral representation for the leading derivative corrections in generic electromagnetic fields at one loop, we specialize to the cases of magnetic- and electric-like field configurations characterized by the vanishing of one of the secular invariants of the electromagnetic field. In these cases, closed-form results and the associated all-orders weak- and strong-field expansions can be worked out.
One immediate application is the leading derivative correction to the renowned Schwinger-formula describing the decay of the quantum vacuum via electron-positron pair production in slowly-varying electric fields.
\end{abstract}

\maketitle

\section{Introduction}\label{sec:intro}

In contrast to the classical notion of vacuum, describing the absence of everything, the vacuum of a quantum field theory (QFT) amounts to a highly non-trivial state.
It is characterized by the omnipresence of quantum fluctuations of all the dynamical degrees of freedom of the underlying QFT.
These fluctuations effectively endow the quantum vacuum with medium-like properties, such as a non-vanishing non-linear response to applied electromagnetic fields.
The latter is in particular triggered by fluctuations of charged particles, which couple directly to electromagnetic fields, and depends on the charges and masses of all fluctuating particles.
Within the Standard Model of particle physics the leading effective interactions between electromagnetic fields are governed by quantum electrodynamics (QED).

A central quantity in
the study of the effective nonlinear interactions of macroscopic electromagnetic fields in the QED vacuum is the Heisenberg-Euler effective action $\Gamma_{\rm HE}$ \cite{Heisenberg:1935qt,Weisskopf:1996bu,Schwinger:1951nm}.
The latter arises from the microscopic theory of QED in a given prescribed (non-quantized) electromagnetic field $\bar F=\bar F^{\mu\nu}$ by integrating out the dynamical degrees of freedom, namely the quantized spinor fields, describing electrons and positrons, and the quantum photon field; cf., e.g., Ref.~\cite{Gies:2016yaa}.
This supplements the classical Maxwell action $\Gamma_{\rm MW}[\bar F]=-\frac{1}{4}\int_x\bar F_{\mu\nu}\bar F^{\mu\nu}$ with effective, nonlinear self-interactions of the prescribed field. 
Apart from the applied electromagnetic field $\bar F$ and derivatives $\partial=\partial^\rho$ thereof, at zero temperature and vanishing chemical potential the only physical parameters characterizing the latter are the electron/positron mass $m$, and the elementary charge $e$ mediating the coupling between charges and electromagnetic fields.
As the quantum fields only appear as virtual states, their momenta are integrated over and hence not determined, eliminating the possibility of any explicit reference to them.
In terms of Feynman diagrams $\Gamma_{\rm HE}[\bar F]$ can be represented as an infinite set of vacuum diagrams, with the charged particle lines dressed to all orders in the external electromagnetic field and its derivatives.
The simplest diagram is a one-loop diagram.
Diagrams featuring more loops are parametrically suppressed with powers of the fine-structure constant $\alpha=e^2/(4\pi)\simeq1/137$.

Upon combination with the speed of light $c$ and the Planck constant $\hbar$, the ratio of $m^2$ and $e$ can be converted into electric $E_{\rm cr}=m^2c^3/(e\hbar)\approx1.3\times10^{18}\,{\rm V}/{\rm m}$ and magnetic $B_{\rm cr}=E_{\rm cr}/c\approx4\times10^9\,{\rm T}$ reference field strengths.
Analogously, the inverse of the electron mass can be converted into spatial $\lambdabar_{\rm C}=\hbar/mc\approx3.8\times10^{-13}\,{\rm m}$ and temporal $\tau_{\rm C}=\lambdabar_{\rm C}/c\simeq1.3\times10^{-21}\,{\rm s}$ reference scales.
The former quantities can be used to render the applied electric and magnetic fields dimensionless, and the latter ones the derivatives.
Hence, is slowly varying electromagnetic fields, characterized by typical spatial (temporal) scales of variation much larger than $\lambdabar_{\rm C}$ ($\tau_{\rm C}$) derivative corrections should be suppressed relatively to contributions scaling with the same power of $\bar F$ but featuring  no derivatives.

The present work is devoted to the study of the leading derivative corrections to the Heisenberg-Euler effective action.
The one-loop Heisenberg-Euler effective action in constant fields has been worked out by Refs.~\cite{Heisenberg:1935qt,Weisskopf:1996bu,Schwinger:1951nm}, an the leading derivative correction by Refs.~\cite{Gusynin:1995bc,Gusynin:1998bt}.
For $\Gamma_{\rm HE}$ in constant fields at two loops, see Refs.~\cite{Ritus:1975cf,Ritus:1977iu,Gies:2016yaa}. Apart from this, higher-loop results in constant fields and lower space-time dimensions \cite{Huet:2011kd,Huet:2018ksz}, as well as one-loop results for specific purely electric or magnetic (one-dimensional) field inhomogeneities are available \cite{Narozhnyi:1970uv,Mamaev:1981dt,Cangemi:1995ee,Dunne:1997kw,Dunne:1998ni,Kim:2009pg}. See also Ref.~\cite{Navarro-Salas:2020oew} for an adiabatic propertime expansion of $\Gamma_{\rm HE}$ at one-loop, and Ref.~\cite{Pegoraro:2021whz} for a study of nonlinear waves in a dispersive vacuum described with a high order derivative electromagnetic Lagrangian.

Our article is organized as follows: after detailing the strategy devised to determine the leading derivative corrections to the Heisenberg-Euler effective action in Sec.~\ref{sec:approach}, we employ our approach to determine the leading derivative correction to the Heisenberg-Euler effective action in Sec.~\ref{sec:calc}. Thereafter, in Sec.~\ref{sec:purefield} we focus on the special cases of magnetic- and  electric-like fields for which only one of the secular invariants of the electromagnetic field does not vanish. Finally, we end with conclusions and an outlook in Sec.~\ref{sec:concls}.

\section{Our Approach}\label{sec:approach}

Here, we demonstrate that the leading derivative correction to the Heisenberg-Euler effective action can efficiently  be determined from the vacuum polarization tensor evaluated in a generic constant and homogeneous background field $\bar F$.
In position space, this correction contains exactly two derivatives but arbitrary powers of the electromagnetic field $\bar F$.
Our derivation -- which is somewhat reminiscent of the approach~\cite{Karbstein:2007be} devised in a different context -- constitutes an alternative route to the result of Gusynin and Shovkovy \cite{Gusynin:1995bc,Gusynin:1998bt}, who determined this correction at one-loop order. 

To this end, we first note that the photon polarization tensor generically mediates a quantum-fluctuation induced effective interaction between two inhomogeneous electromagnetic fields characterized by the vector potential $A(x)$.
In turn, it is a central ingredient to the effective action describing the physics of arbitrary-frequency fields in the presence of a constant background field \cite{Gies:1999vb,Karbstein:2011ja}. In position space, this effective action reads
\begin{equation}
 \Gamma[A(x)]=-\frac{1}{4}\int_xF_{\mu\nu}(x)F^{\mu\nu}(x)-\frac{1}{2} \int_x\int_{x'} A_\mu(x)\Pi^{\mu\nu}(x-x'|\bar F)A_\nu(x') + {\cal O}(A^3)\,. \label{eq:GammaAgenfreq}
\end{equation}
Here, $F(x)$ denotes the field strength tensor of the manifestly inhomogeneous field $A(x)$, and $\Pi^{\mu\nu}(x-x'|\bar F)$ is the polarization tensor evaluated in the background field $\bar F$.
The neglected higher-order terms encode effective self-interactions of the field $A(x)$. To keep notations compact, throughout this work we employ the shorthand notations $\int_x\equiv\int{\rm d}^4x$ and $\int_k\equiv\int{\rm d}^4k/(2\pi)^4$ for integrations over position and momentum space, respectively. Moreover, we use the Heaviside-Lorentz System with $c=\hbar=1$; 
$g^{\mu\nu}={\rm diag}(-1,+1,+1,+1)$.

Due to translational invariance in homogeneous constant fields, in momentum space the polarization tensor $\Pi^{\mu\nu}(k,k'|\bar F)=\int_x \int_{x'}{\rm e}^{{\rm i}kx}\,\Pi^{\mu\nu}(x-x'|\bar F)\,{\rm e}^{{\rm i}k'x'}$ does not depend explicitly on both the in- and outgoing momenta, but is a function of the momentum transfer $k$ only. This implies that $\Pi^{\mu\nu}(k,k'|\bar F)\sim(2\pi)^4\delta(k+k')$ and resembles the situation at zero background field, where the vacuum polarization tensor can be solely expressed in terms of $k$.
There, the Ward identity $k_\mu\Pi^{\mu\nu}=\Pi^{\mu\nu}k_\nu=0$ immediately constrains its tensor structure to be spanned by $(k^2g^{\mu\nu}-k^{\mu}k^{\nu})$.
In the present case, the field strength tensor of the background field $\bar F$ provides an additional building block to form tensor structures compatible with the Ward identity.
However, as both $\Pi$ and $\bar F$ have two Minkowski indices, and the former is a function of $k$ and $\bar F$ only, $\Pi^{\mu\nu}(k,k'|\bar F)$ has to be even in $k$.
Besides, it is even in $\bar F$ and regular at $\bar F=0$.

Upon transformation to position space, insertion into \Eqref{eq:GammaAgenfreq}, and making use of partial integrations, the contribution to $\Pi^{\mu\nu}(k,k'|\bar F)$ which is quartic in $k$  gives rise to an effective interaction term  which can be schematically expressed as $\Gamma[F,\bar F]|_{\sim\partial^2}=\int_x\,h^{(2)}(\bar F)\,(\partial F)^2$.
Here, the scalar function $h^{(2)}(\bar F)$ accounts for arbitrary powers of the background field $\bar F$, and we explicitly ensured that a single derivative acts on each factor of the inhomogeneous fields $F(x)$.
Finally, substituting $\partial F\to \partial\bar F$, where $\bar F=\bar F(x)$ is now to be understood as slowly varying electromagnetic field, we arrive at
\begin{equation}
 \Gamma_{\rm HE}[\bar F]|_{\sim\partial^2}=\int_x\,h^{(2)}(\bar F)\,(\partial\bar F)^2\,, \label{eq:GammaPartial2}
\end{equation}
which corresponds to the desired derivative correction to the Heisenberg-Euler effective action $\Gamma_{\rm HE}$ featuring exactly two derivatives, but arbitrary powers of the slowly varying field.

As the derivation of $\Pi^{\mu\nu}(k,k'|\bar F)$ explicitly accounts for all possible variants of coupling the in- and out-fields with momenta $k$, $k'$ and Minkowski indices $\mu$, $\nu$ to the charged particle loop, the procedure outlined above indeed ensures that \Eqref{eq:GammaPartial2} can be identified with the leading derivative correction to the Heisenberg-Euler effective action in the field $F=\bar F+\partial\bar F+\ldots$
We emphasize that for this identification the regrouping of the terms such that each power of the inhomogeneous field $F$ comes with a derivative acting on it prior to the substitution is absolutely essential. 

Moreover, we note that though upon insertion into \Eqref{eq:GammaAgenfreq} and appropriate integrations by parts, the contribution to $\Pi^{\mu\nu}(k,k'|\bar F)$ which is quadratic in $k$ results in a contribution $\sim \int_x\,h^{(0)}(\bar F)\,F^2$, this expression does not reproduce the zero-derivative result for $\Gamma_{\rm HE}$ in the limit of $F\to \bar F$. The reason for this is the fact that in the derivation of the photon polarization tensor and \Eqref{eq:GammaAgenfreq} the fields $\bar F$ and $F$ are assumed to be manifestly different. Inconsistencies arise as soon as (at least) one of the couplings to the field $F$ is identified with a coupling to the background field.

The contributions to $\Pi^{\mu\nu}(k,k'|\bar F)$ beyond quartic order, which translate into higher-order $n$ derivative terms are also not helpful for the purpose of a systematic derivation of higher-order derivative corrections to $\Gamma_{\rm HE}$.
This is a direct consequence of the fact that there is no unambiguous way in assigning the additional derivatives to any of the two inhomogeneous fields $F(x)$ before invoking the substitution $F\to \bar F$. The possibility of partial integrations, which after this substitution also act on the factors of $\bar F$ in the scalar functions $h^{(2n)}(\bar F)$, renders different assignments inequivalent for $n>1$, and imply inconsistent results.

On the other hand, along the lines outlined above the result for the contribution to $\Gamma_{\rm HE}$ containing $n$ derivatives, but arbitrary powers of the field could be extracted from the $n$-rank polarization tensor evaluated in the homogeneous constant background field $\bar F$.
As the determination of the $n$-derivative contribution only requires knowledge of the term scaling as $k^{2n}\sim k^{\sigma_1}\ldots k^{\sigma_{2n}}$ of the $n$-rank polarization tensor, aiming at the evaluation of the respective contribution in cases where the required polarization tensor has not yet been determined, for this endeavor it suffices to determine this tensor only at an accuracy of order $k^{2n}$.

\section{Explicit Calculation}\label{sec:calc}

Subsequently, we employ the strategy outlined above to explicitly determine the quadratic derivative correction to the Heisenberg-Euler effective action at one loop \cite{Gusynin:1995bc,Gusynin:1998bt}.
The determination of this contribution is particularly straightforward because $\Pi^{\mu\nu}(k,k'|\bar F)$ is known analytically at one-loop order \cite{BatShab,Baier:1974hn,Urrutia:1977xb,Dittrich:2000wz,Schubert:2000yt,Dittrich:2000zu}.
However, we emphasize that our approach is not limited to one loop. For instance, a result for the two-loop photon polarization tensor evaluated in a homogeneous constant background field could be readily employed to extract the quadratic derivative correction to $\Gamma_{\rm HE}$ at two loops.

Following the notations of \cite{Dittrich:2000zu}, the photon polarization tensor can be expressed as
\begin{equation}
    \Pi^{\mu\nu}(k,k'|\bar F)=(2\pi)^4\delta(k+k')\Bigl\{\Pi_0 P_T^{\mu\nu} +(\Pi_\perp-\Pi_0) P_\perp^{\mu\nu} +(\Pi_\parallel-\Pi_0) P_\parallel^{\mu\nu} +\pi_Q Q^{\mu\nu}\Bigr\}\,, \label{eq:PiF0}
\end{equation}
where
$\Pi_{0,\parallel,\perp}$ and $\pi_Q$ are scalar functions which depend both on the background field $\bar F$ and the transferred momentum $k$. Its tensor structure is spanned by 
\begin{equation}
    P^{\mu\nu}_T=g^{\mu\nu}-\frac{k^\mu k^\nu}{k^2}\,, \quad
    P^{\mu\nu}_\perp=\frac{v_\perp^\mu v_\perp^\nu}{v_\perp^2}\,, \quad
    P^{\mu\nu}_\parallel=\frac{v_\parallel^\mu v_\parallel^\nu}{v_\parallel^2}\,, \quad
    Q^{\mu\nu}=v_\parallel^\mu v_\perp^\nu +v_\perp^\mu v_\parallel^\nu\,, \label{eq:Tensors}
\end{equation}
where the four-vectors $v_{\parallel,\perp}$ are defined as 
\begin{equation}
 v_{\parallel/\perp}^\mu=\frac{c_{\pm}(k{}^\star\!\bar F)^\mu\mp c_\mp (k\bar F)^\mu}{c_+^2+c_-^2}\,, \quad\text{such that}\quad v_{\parallel/\perp}^2=\frac{(k\bar F)^2\mp k^2c_\pm^2}{c_+^2+c_-^2}\,.
\end{equation}
Also note that $v_\perp^2-v_\parallel^2=k^2$.
Here, we use the shorthand notation $(k\bar F)^\mu=k_\nu\bar F^{\nu\mu}$, etc., and $c_\pm$ denote the secular invariants of the electromagnetic field. The latter are related to the gauge and Lorentz invariants ${\cal F}=\frac{1}{4}\bar F_{\mu\nu}\bar F^{\mu\nu}$ and ${\cal G}=\frac{1}{4}\bar F_{\mu\nu}{}^\star\!\bar F^{\mu\nu}$ as $c_\pm=(\sqrt{{\cal F}^2+{\cal G}^2}\pm{\cal F})^{1/2}$; ${}^\star\!\bar F^{\mu\nu}$ is the dual field strength tensor.
The above definitions are such that the three tensors $P_{\parallel,\perp}^{\mu\nu}$ and $P_0^{\mu\nu}=P_T^{\mu\nu}-P_\parallel^{\mu\nu}-P_\perp^{\mu\nu}$ are projectors and fulfill the usual projector identities. At the same time, $Q^{\mu\nu}$ is only orthogonal to $P_0^{\mu\nu}$ and not a projector.

Defining $\pi_T=\Pi_0/k^2$ and $\pi_{\parallel/\perp}=(\Pi_{\parallel/\perp}-\Pi_0)/v_{\parallel/\perp}^2$, \Eqref{eq:PiF0} can alternatively
be represented as
\begin{align}
 \Pi^{\mu\nu}(k,k'|\bar F)&=(2\pi)^4\delta(k+k')\Bigl\{(k^2g^{\mu\nu}-k^\mu k^\nu)\,\pi_T + (k\bar F)^\mu(k\bar F)^\nu \pi_{\bar F\bar F} + (k{}^\star\!\bar F)^\mu(k{}^\star\!\bar F)^\nu\pi_{{}^\star\!\bar F{}^\star\!\bar F} \nonumber\\
 &\hspace*{3.5cm}+[(k{}^\star\!\bar F)^\mu(k\bar F)^\nu + (k\bar F)^\mu(k{}^\star\!\bar F)^\nu]\pi_{{}^\star\!\bar F \bar F}\Bigr\}\,.
 \label{eq:Pi_F}
\end{align}
The scalar coefficients $\pi_p$ in \Eqref{eq:Pi_F} are given by
\begin{align}
 \pi_{\bar F \bar F}&=\frac{1}{(c_+^2+c_-^2)^2}\bigl[c_+^2\pi_\perp+c_-^2\pi_\parallel-2c_+c_-\pi_Q\bigr]\,, \nonumber\\
 \pi_{{}^\star\!\bar F{}^\star\!\bar F}&=\frac{1}{(c_+^2+c_-^2)^2}\bigl[c_-^2\pi_\perp+c_+^2\pi_\parallel+2c_+c_-\pi_Q\bigr]\,, \nonumber\\
 \pi_{{}^\star\!\bar F\bar F}&=\frac{1}{(c_+^2+c_-^2)^2}\bigl[c_+c_-(\pi_\perp-\pi_\parallel)+(c_+^2-c_-^2)\pi_Q\bigr]\,.
 \label{eq:pi_FF}
\end{align}
While this structure is general, the explicit expressions for the scalar functions encoding the nontrivial dependences on $\bar F$ and $k$ at one loop order can be cast in the following form,
\begin{equation}
 \left\{\begin{array}{c}
         \!\!\pi_T\!\! \\ \!\!\pi_\parallel\!\! \\ \!\!\pi_\perp\!\! \\ \!\!\pi_Q\!\! \\
        \end{array}\right\}
 =\frac{\alpha}{2\pi}\int_0^\infty\frac{{\rm d}s}{s}\,{\rm e}^{-{\rm i}m^2s}\!\left[\int_0^1{\rm d}\nu\,{\rm e}^{-{\rm i}(v_\perp^2 n_\perp-v_\parallel^2n_\parallel)s}\,\frac{zz'}{\sin z\sinh z'}
 \left\{\begin{array}{c}
         N_0 \\ \!\!N_0-N_1\!\! \\ \!\!N_2-N_0\!\! \\ -N_3
        \end{array}\right\}
 - \left\{\begin{array}{c}
         \!\!\frac{2}{3}\!\! \\ 0 \\ 0 \\ 0
        \end{array}\right\}\right]\!, \label{eq:pi}
\end{equation}
with
\begin{align}
 N_0&=\cos(\nu z)\cosh(\nu z')-\cot z\sin(\nu z)\coth z'\sinh(\nu z')\,, \nonumber\\
 N_1&=2\cos z\,\frac{\cosh z'-\cosh(\nu z')}{\sinh^2 z'}\,, \quad N_2=N_1|_{z\leftrightarrow-{\rm i}z'}\,, \nonumber\\
 N_3&=\frac{1-\cos z\cos(\nu z)}{\sin z}\frac{1-\cosh z'\cosh(\nu z')}{\sinh z'}+\sin(\nu z)\sinh(\nu z')\,, \nonumber\\
 n_\parallel&=\frac{\cosh z'-\cosh(\nu z')}{2z'\sinh z'}\,, \quad n_\perp=n_\parallel|_{z\leftrightarrow-{\rm i}z'}\,,
 \end{align}
where we used the shorthand notations $z=ec_+s$ and $z'=ec_-s$. Here and in the following, the prescription $m^2\to m^2-{\rm i}0^+$ for the square of the electron mass $m$ is implicitly assumed. Besides, the integration contour of the propertime integration is implicitly assumed to lie slightly below the real positive axis \cite{Karbstein:2013ufa}.

Note, that the entire momentum dependence of \Eqref{eq:pi} is encoded in the phase of the propertime integral over $s$.
Hence, it is obvious that all the scalar functions $\pi_p$ introduced above can be formally expanded as $\pi_p=\sum_{n=0}^\infty\pi_p^{(2n)}$, with $\pi_p^{(2n)}\sim k^{2n}$.
The contributions $\pi_p^{(2n)}$ constitute the photon polarization tensor $\Pi^{\mu\nu}(k,k'|\bar F)$ at order $k^{2n+2}$. In turn, here we are specifically interested in $\pi_p^{(2)}$; the polarization at this order $\Pi^{(2)\mu\nu}(k,k'|\bar F)$ follows from \Eqref{eq:Pi_F} upon substitution of the coefficients $\pi_p\to\pi_p^{(2)}$.

Clearly, central building blocks to $\pi_p^{(2)}$ are
\begin{equation}
{\cal N}_i^{\parallel}=\int_0^1{\rm d}\nu\,n_{\parallel}N_i \quad\text{and}\quad {\cal N}_i^{\perp}=\int_0^1{\rm d}\nu\,n_{\perp}N_i\,, \label{eq:Niparallelperp}
\end{equation}
with $i\in\{0,1,2,3\}$.
The integral over $\nu$ in \Eqref{eq:Niparallelperp} can be performed explicitly, yielding
\begin{align}
 {\cal N}^\parallel_0(z,z')&=\frac{1}{z^2+4z'^2}\frac{z'}{z}\biggl[\frac{3}{2}\frac{z^2}{z^2+z'^2}\coth z'\Bigl(\frac{\cosh z'}{\sin z}-\frac{z'}{z}\frac{\cos z}{\sinh z'}\Bigr)-\frac{\sin z}{\sinh z'}\biggr], \nonumber\\
 {\cal N}^\parallel_1(z,z')&=\frac{1}{z'}\frac{\cos z}{\sinh z'}\biggl[1+\frac{3}{2}\frac{1}{\sinh z'}\Bigl(\frac{1}{\sinh z'}-\frac{\cosh z'}{z'}\Bigr)\biggr], \nonumber\\
 {\cal N}_2^\parallel(z,z')&=\frac{\cosh z'}{\sin z}\biggl[\frac{1}{z^2+z'^2}\Bigl(\frac{z'}{z}\coth z' +\frac{z^2}{z'^2}\cot z\Bigr)-\frac{\cot z\coth z'}{z'}\biggr], \nonumber\\
 {\cal N}^\parallel_3(z,z')&=\frac{3}{4}\frac{\coth z'}{\sin z}\frac{1}{z'}\Bigl(\frac{1}{\sinh z'} -\frac{\cosh z'}{z'}\Bigr)+\frac{1}{2}\frac{1}{z^2+z'^2}\frac{z^2}{z'^2}\frac{\sinh z'}{\sin z} \nonumber\\
 &\quad+\frac{3}{2}\frac{1}{z^2+4z'^2}\frac{z'^2}{z^2+z'^2}\biggl[2\coth z'\Bigl(\frac{\cosh z'}{\sin z}-\frac{z'}{z}\frac{\cos z}{\sinh z'}\Bigr)-\frac{\sin z}{\sinh z'}\biggr],
 \label{eq:calN}
\end{align}
as well as ${\cal N}_0^\perp={\cal N}_0^\parallel|_{z\leftrightarrow-{\rm i}z'}$, $ {\cal N}^\perp_1={\cal N}^\parallel_2 |_{z\leftrightarrow-{\rm i}z'}$, $ {\cal N}^\perp_2={\cal N}^\parallel_1 |_{z\leftrightarrow-{\rm i}z'}$ and $ {\cal N}^\perp_3={\cal N}^\parallel_3 |_{z\leftrightarrow-{\rm i}z'}$.
For completeness and later reference, we also provide the leading contributions of these quantities in a weak-field expansion. The respective results are
\begin{align}
 {\cal N}^\parallel_0(z,z')&=\frac{2}{15}-\frac{e^2}{315}(c_+^2+2c_-^2)s^2+{\cal O}(\bar F^4)\,, \nonumber\\
 {\cal N}^\parallel_1(z,z')&=\frac{2}{15}-\frac{e^2}{315}(21c_+^2+13c_-^2)s^2+{\cal O}(\bar F^4)\,, \nonumber\\
 {\cal N}^\parallel_2(z,z')&=\frac{2}{15}+\frac{e^2}{315}(10c_+^2+18c_-^2)s^2+{\cal O}(\bar F^4)\,, \nonumber\\
 {\cal N}^\parallel_3(z,z')&=-\frac{e^2}{35}c_+c_-s^2+{\cal O}(\bar F^4)\,.  \label{eq:calN_series}
 \intertext{Moreover, note that}
 \frac{zz'}{\sin z\sinh z'}&=1+\frac{e^2}{6}(c_+^2-c_-^2)s^2+{\cal O}(\bar F^4)\,. \nonumber
\end{align}

Introducing the shorthand notations
\begin{equation}
 {\cal N}_i^-={\cal N}_i^\parallel-{\cal N}_i^\perp \quad\text{and}\quad {\cal N}_i^+=\frac{c_+^2}{c_+^2+c_-^2}\,{\cal N}_i^\parallel+\frac{c_-^2}{c_+^2+c_-^2}\,{\cal N}_i^\perp\,,
\end{equation}
the functions $\pi_p^{(2)}$ can then be compactly expressed as
\begin{equation}
\pi^{(2)}_p = k_\alpha h^{\alpha\beta}_p(\bar F)k_\beta\,,
\quad\text{with}\quad
 h^{\alpha\beta}_p(\bar F) = \frac{\bar F^\alpha_{\ \, \tau}\bar F^{\beta\tau}}{c_+^2+c_-^2} h_p^-(c_+,c_-) -g^{\alpha\beta}h_p^+(c_+,c_-)\,,
 \label{eq:h_p}
\end{equation}
where
\begin{equation}
  \left\{\begin{array}{c}
         h_T^\pm \\ \!h_\parallel^\pm\! \\ \!h_\perp^\pm\! \\ h_Q^\pm
        \end{array}\right\}
={\rm i}\frac{\alpha}{2\pi}\int_0^\infty{\rm d}s\,{\rm e}^{-{\rm i}m^2s}\,\frac{zz'}{\sin z\sinh z'}
 \left\{\begin{array}{c}
         {\cal N}_0^\pm \\ \!{\cal N}_0^\pm-{\cal N}_1^\pm\! \\ \!{\cal N}_2^\pm-{\cal N}_0^\pm\! \\ -{\cal N}_3^\pm
        \end{array}\right\}.
 \label{eq:hs}
\end{equation}
The fact that the functions $\pi^{(2)}_p$ are regular at $c_+=c_-=0$ and feature asymptotic expansions in terms of combinations of $c_+$ and $c_-$ is not obvious.
However, at least at low orders one can easily convince oneself that this is indeed the case by performing explicit expansions; cf. also \Eqref{eq:calN_series}.
Besides, we note that Eqs.~\eqref{eq:pi_FF} and \eqref{eq:h_p} imply that $h_{\bar F\bar F}^{\alpha\beta}$ ($h_{\bar F\bar F}^\pm$) relates to $h_\perp^{\alpha\beta}$, $h_\parallel^{\alpha\beta}$ and $h_Q^{\alpha\beta}$ ($h_\perp^\pm$, $h_\parallel^\pm$ and $h_Q^\pm$) in exactly the same way as $\pi_{\bar F\bar F}$ relates to $\pi_\perp$, $\pi_\parallel$ and $\pi_Q$, etc.

With these preparations, we can now explicitly determine the quadratic derivative correction $\Gamma_{\rm HE}[\bar F]|_{\sim \partial^2}$ to the Heisenberg-Euler effective action.
Following the strategy outlined above and using the Fourier representation of the gauge field $A^\mu(x)=\int_k{\rm e}^{{\rm i}kx}A^\mu(k)$, we first evaluate the quantity 
\begin{equation}
 \Gamma[F,\bar F]\big|_{\sim\partial^2} =-\frac{1}{2}\int_k \int_{k'} A_\mu(k)\,\Pi^{(2)\mu\nu}(k,k'|\bar F)\, A_\nu(k')\,.
 \label{eq:DeltaGamma}
\end{equation}
A direct consequence of our definition of the momentum space representation of the gauge field is
$F^{\mu\nu}(x)= \int_k{\rm e}^{{\rm i}kx} F^{\mu\nu}(k)$ with $F^{\mu\nu}(k)={\rm i}\bigl(k^\mu A^\nu(k)-k^\nu A^\mu(k)\bigr)$.
Therewith it is easy to show that $(k\bar F)^\nu A_\nu(k)=-{\rm i}\bar{F}^{\rho\nu} F_{\rho\nu}(k)/2$ and analogously $(k {}^*\!\bar F)^\nu A_\nu(k)=-{\rm i}\,{}^*\!\bar{F}^{\rho\nu} F_{\rho\nu}(k)/2$.
Correspondingly, we find
\begin{align}
 \Gamma[F,\bar F]\big|_{\sim\partial^2}&=\frac{1}{4}\int_k \,\Bigl\{\bigl[k_\alpha F_{\mu\nu}(k)\bigr]\!\bigl[-k_\beta F^{\mu\nu}(-k)\bigr]\, h^{\alpha\beta}_T(\bar F)\nonumber\\
 & \hspace*{2cm} + \frac{1}{2}\bigl[ k_\alpha F_{\sigma\mu}(k)\bigr]\!\bigl[- k_\beta F_{\rho\nu}(-k)\bigr] \Bigl[\bar{F}^{\sigma\mu}\bar{F}^{\rho\nu} h^{\alpha\beta}_{\bar F\bar F}(\bar F) + {}^*\!\bar{F}^{\sigma\mu} {}^*\!\bar{F}^{\rho\nu} h^{\alpha\beta}_{{}^*\! \bar F{}^*\! \bar F}(\bar F)\nonumber\\
 &\hspace*{8cm} + 2 {}^*\!\bar{F}^{\sigma\mu}\bar{F}^{\rho\nu} h^{\alpha\beta}_{{}^*\! \bar F \bar F}(\bar F)\Bigr]\Bigr\}\,.
 \label{eq:DeltaGamma2}
\end{align}
Accounting for the identities $\int_k {\rm e}^{{\rm i}kx} [k_\alpha F_{\mu\nu}(k)]=-{\rm i}\partial_\alpha F_{\mu\nu}(x)$ and $\int_k u(k)v(-k)=\int_x u(x)v(x)$, this expression can be readily transformed to position space.

Finally substituting $F\to\bar F(x)$ and $\bar F\to\bar F(x)$, \Eqref{eq:DeltaGamma2} yields the desired contribution to the Heisenberg-Euler effective action,
\begin{align}
 \Gamma_{\rm HE}[\bar F]\big|_{\sim \partial^2}&=-\frac{1}{4}\int_x \,\Bigl\{\partial_\alpha \bar F_{\mu\nu} \partial_\beta \bar F^{\mu\nu}\, h^{\alpha\beta}_T(\bar F)\nonumber\\
 & \hspace*{2.2cm} + \frac{1}{2} \partial_\alpha \bar F_{\sigma\mu} \partial_\beta \bar F_{\rho\nu} \Bigl[\bar{F}^{\sigma\mu}\bar{F}^{\rho\nu} h^{\alpha\beta}_{\bar F\bar F}(\bar F) + {}^*\!\bar{F}^{\sigma\mu} {}^*\!\bar{F}^{\rho\nu} h^{\alpha\beta}_{{}^*\!\bar F{}^*\!\bar F}(\bar F)\nonumber\\
 &\hspace*{7cm} + 2 {}^*\!\bar{F}^{\sigma\mu}\bar{F}^{\rho\nu} h^{\alpha\beta}_{{}^*\!\bar F \bar F}(\bar F)\Bigr]\Bigr\}\,,
 \label{eq:Gamma2derivs_gen}
\end{align}
where $\bar F=\bar F(x)$ is to be implicitly understood. We note that this expression with tensor structures~\eqref{eq:h_p} is generic and holds at all loop orders.

With the help of \Eqref{eq:calN_series}, we infer the following weak-field limits for the tensor structures in \Eqref{eq:Gamma2derivs_gen} at one loop,
\begin{align}
 h_T^{\alpha\beta}(\bar F)&=-\frac{1}{15}\frac{\alpha}{\pi}\frac{1}{m^2}\Bigl[1-\frac{1}{7}\Bigl(\frac{e}{m^2}\Bigr)^2\bar F_{\kappa\lambda}\bar F^{\kappa\lambda}\Bigr]g^{\alpha\beta}+\frac{1}{105}\frac{\alpha}{\pi}\frac{1}{m^2}\Bigl(\frac{e}{m^2}\Bigr)^2 \bar F^\alpha_{\ \, \tau}\bar F^{\beta\tau}+{\cal O}(\bar F^4)\,, \nonumber\\
 h_{\bar F\bar F}^{\alpha\beta}(\bar F)&=\frac{11}{315}\frac{\alpha}{\pi}\frac{1}{m^2}\Bigl(\frac{e}{m^2}\Bigr)^2g^{\alpha\beta}+{\cal O}(\bar F^2)\,, \nonumber\\
 h_{{}^*\!\bar F{}^*\!\bar F}^{\alpha\beta}(\bar F)&=\frac{4}{63}\frac{\alpha}{\pi}\frac{1}{m^2}\Bigl(\frac{e}{m^2}\Bigr)^2g^{\alpha\beta}+{\cal O}(\bar F^2)\,, \nonumber\\
 h_{{}^*\!\bar F\bar F}^{\alpha\beta}(\bar F)&={\cal O}(\bar F^2)\,. \label{eq:weakfield}
\end{align}
Upon plugging these results into \Eqref{eq:Gamma2derivs_gen} and using the identity~\eqref{eq:id:*F*FFF} to eliminate the dependences of the dual field strength tensor, we obtain
\begin{align}
 {\cal L}_{\rm HE}^{1\text{-loop}}(\bar F)\big|_{\sim \partial^2}&=\frac{1}{60}\frac{\alpha}{\pi}\frac{1}{m^2}\partial_\alpha\bar F_{\mu\nu}\partial^\alpha\bar F^{\mu\nu} \nonumber\\
 &\quad+\frac{\alpha}{\pi}\Bigl(\frac{e}{m^2}\Bigr)^2\frac{1}{m^2}\Bigl[\frac{1}{180}\bar F_{\mu\nu}\bar F^{\mu\nu}\partial^\alpha\bar F_{\rho\sigma}\partial_\alpha\bar F^{\rho\sigma} +\frac{1}{280}\bar F_{\mu\nu}\bar F_{\rho\sigma}\partial^\alpha \bar F^{\mu\nu} \partial_\alpha\bar F^{\rho\sigma} \nonumber\\
 &\hspace*{2cm}-\frac{2}{63}\bar F_{\rho\mu}\bar F^{\sigma\mu} \partial^\alpha\bar F_{\sigma\nu} \partial_\alpha\bar F^{\rho\nu}-\frac{1}{420}\bar F_{\rho\sigma}\bar F^{\rho\alpha}\partial^\sigma\bar F_{\mu\nu}\partial_\alpha\bar F^{\mu\nu}\Bigr] \nonumber\\
 &\quad+{\cal O}(\bar F^6)\,. \label{eq:LHEpartial^2LO}
\end{align}
It is noteworthy that the contribution to \Eqref{eq:LHEpartial^2LO} which is quartic in the field strength can be expressed in terms of just four different tensor structures.

\section{Magnetic- and Electric-like Field Configurations}\label{sec:purefield}

In the remainder, we focus on the special situation where only one of the two invariants $c_+$ or $c_-$ does not vanish. The remaining parameter may be arbitrarily strong.
This grants access to the cases of a purely magnetic and electric field, respectively.
In this case additional insights are possible and (i) the asymptotic expansion for perturbatively weak fields can be organized in terms of a single infinite sum, with all the expansion coefficients known explicitly. Besides, (ii) the propertime integration over $s$ can even be performed explicitly and the result can be expressed in terms of the Hurwitz zeta function $\zeta(l,\chi)=\sum_{n=0}^\infty(\chi+n)^{-l}$ and derivatives thereof; primes on $\zeta$ denote derivatives with respect to $l$.

First of all, we note that for either $c_+=0$ or $c_-=0$ \Eqref{eq:Gamma2derivs_gen} simplifies significantly due to the fact that in this case $\partial_\alpha\bar F_{\sigma\mu}{}^*\!\bar F^{\sigma\mu}=2\partial_\alpha{\cal G}=0$, which implies that
\begin{align}
 {\cal L}_{\rm HE}(\bar F)\big|_{\sim \partial^2}=-\frac{1}{4}\partial_\alpha\bar F_{\mu\nu} \partial_\beta\bar F^{\mu\nu}\, h^{\alpha\beta}_T(\bar F)
 - \frac{1}{8} \partial_\alpha\bar F_{\sigma\mu} \partial_\beta\bar F_{\rho\nu}\bar{F}^{\sigma\mu}\bar{F}^{\rho\nu} h^{\alpha\beta}_{\bar F\bar F}(\bar F) \,.
 \label{eq:Gamma2derivs_B(E)}
\end{align}
Hence, the only quantities to be determined in this specific limit are $h^{\alpha\beta}_T(F)$ and $h^{\alpha\beta}_{F F}(F)$.
Aiming at their explicit determination, we note that for finite $c_+$ but $c_-=0\leftrightarrow z'=0$ we have
\begin{align}
 \frac{z}{\sin z}{\cal N}^\parallel_0(z,0)&=-\frac{1}{z^2}\Bigl[\frac{3}{2}\Bigl(\partial_z+\frac{1}{z}\Bigr)\cot z+1\Bigr], \nonumber\\
 \frac{z}{\sin z}{\cal N}^\parallel_1(z,0)&=\frac{2}{15}z\cot z\,, \nonumber\\
 \frac{z}{\sin z}{\cal N}^\parallel_2(z,0)&=-\Bigl[\frac{1}{2}\Bigl(\frac{1}{z}+\frac{z}{3}\Bigr)\partial_z+\frac{1}{z^2}\Bigr]\partial_z\cot z, \nonumber\\
 \frac{z}{\sin z}{\cal N}^\parallel_3(z,0)&=0\,,
 \label{eq:pure_z_parallel}
\end{align}
and
\begin{align}
 \frac{z}{\sin z}{\cal N}^\perp_0(z,0)&=-\frac{3}{8}\Bigl\{\frac{1}{z^2}+\Bigl[\frac{1}{2z}\partial_z+\Bigl(\frac{1}{z^2}+\frac{2}{3}\Bigr)\Bigr]\partial_z\cot z\Bigr\}\,, \nonumber\\
 \frac{z}{\sin z}{\cal N}^\perp_1(z,0)&=\Bigl[\Bigl(\frac{1}{z^2}+\frac{1}{3}\Bigr)\partial_z+\frac{1}{z^3}\Bigr]\cot z+\frac{1}{z^2}+\frac{1}{3}\,, \nonumber\\
 \frac{z}{\sin z}{\cal N}_2^\perp(z,0)&=-\frac{1}{4}\Bigl(\partial_z+\frac{3}{z}\Bigr)\partial_z^2\cot z \nonumber\\
 \frac{z}{\sin z}{\cal N}^\perp_3(z,0)&=0\,.
 \label{eq:pure_z_perp}
\end{align}
Obviously, these quantities be written entirely in terms of products of powers of $z$ and $\cot z$ as well as derivatives thereof.

The analogous expressions for $c_+=0\leftrightarrow z=0$ but finite $c_-$ follow straightforwardly with the identities given below \Eqref{eq:calN}.
In turn, the only two non-trivial identities needed to determine the perturbative weak field expansions of \Eqref{eq:Gamma2derivs_B(E)} are
\begin{equation}
 \cot(z)=\sum_{n=0}^\infty(-1)^n\frac{2^{2n}{\cal B}_{2n}}{(2n)!}z^{2n-1}\quad\text{for}\quad |z|<\pi\,,\quad\quad
(\text{\cite{Gradshteyn}}: 1.411.11)
\end{equation}
where ${\cal B}_{2n}$ denote Bernoulli numbers, and
\begin{equation}
 \int_0^\infty{\rm d}s\,z^{n+\epsilon}\,{\rm e}^{-{\rm i}m^2 s} =\frac{1}{ec_+}\frac{\Gamma(n+1+\epsilon)}{{\rm i}^{n+1+\epsilon}}\Bigl(\frac{ec_+}{m^2}\Bigr)^{n+1+\epsilon} \,,\quad\quad
 (\text{\cite{Gradshteyn}}: 3.551.2)\label{eq:ints1}
\end{equation}
which holds individually for $n+\epsilon>-1$.
Therewith we infer the following expressions for the scalar coefficients determining the tensors $h_p^{\alpha\beta}$ in \Eqref{eq:Gamma2derivs_B(E)} for the case of $c_-=0$,
\begin{align}
 h_T^+(c_+,0)&=-\frac{\alpha}{\pi}\frac{1}{m^2}\sum_{n=0}^\infty\frac{12{\cal B}_{2(n+2)}}{(2n+1)(2n+2)(2n+3)}\Bigl(\frac{2ec_+}{m^2}\Bigr)^{2n}, \nonumber\\
 h_T^-(c_+,0)&=\frac{\alpha}{\pi}\frac{1}{m^2}\sum_{n=1}^\infty \frac{1}{4}\frac{1}{n+1}\Bigl[\frac{3(2n-5){\cal B}_{2(n+2)}}{(2n+1)(2n+3)}-{\cal B}_{2(n+1)}\Bigr]\Bigl(\frac{2ec_+}{m^2}\Bigr)^{2n}\,, \label{eq:h+-B}\\
 h^+_{\bar F\bar F}(c_+,0)&=-\frac{\alpha}{\pi}\frac{1}{m^2}\Bigl(\frac{e}{m^2}\Bigr)^2 \sum_{n=0}^\infty 4\frac{n+1}{n+2}\Bigl[\frac{4{\cal B}_{2(n+3)}}{(2n+3)(2n+5)}-\frac{{\cal B}_{2(n+2)}}{3}\Bigr]\Bigl(\frac{2ec_+}{m^2}\Bigr)^{2n}\,, \nonumber\\
 h^-_{\bar F\bar F} (c_+,0)&=\frac{\alpha}{\pi}\frac{1}{m^2}\Bigl(\frac{e}{m^2}\Bigr)^2\sum_{n=1}^\infty\frac{1}{n+2}\Bigl[\frac{16n^2+50n+49}{(2n+3)(2n+5)}{\cal B}_{2(n+3)}+\frac{4n+7}{3}{\cal B}_{2(n+2)}\Bigr]\Bigl(\frac{2ec_+}{m^2}\Bigr)^{2n}\,. \nonumber
\end{align}

On the other hand, when aiming at performing the propertime integration over $s$ without resorting to an expansion, we need another identity apart from \Eqref{eq:ints1}, namely
\begin{align}
 &\int_0^\infty{\rm d}s\,(as)^{n+\epsilon}\,{\rm e}^{-{\rm i}m^2 s}\, {\rm coth}(as) \nonumber\\
 &\quad\quad=\frac{1}{a}\frac{\Gamma(n+1+\epsilon)}{2^{n+1+\epsilon}}\biggl[2\zeta\bigl(n+1+\epsilon,\tfrac{{\rm i}m^2}{2a}\bigr)-\Bigl(\frac{2a}{{\rm i}m^2}\Bigr)^{n+1+\epsilon}\biggr] \,,  \quad\quad
 (\text{\cite{Gradshteyn}}: 3.551.3)\label{eq:ints2}
\end{align}
which holds individually for $n+\epsilon>0$ and $a=|a|\,{\rm e}^{{\rm i}\delta}$ with $0\leq\delta<\frac{\pi}{2}$. The conditions on $n+\epsilon$ are rendered irrelevant upon combination of these integrals in the explicit determination of the coefficients $h_p^\pm$.
To perform the integrals involving derivatives of $\cot z$ we moreover make use of the identity $\partial_z^n \cot z=\frac{1}{z^n}\partial_c^n\,\cot(cz)\big|_{c=1}$.
The resulting expressions for the coefficients encoding the non-trivial field dependence of \Eqref{eq:Gamma2derivs_B(E)} in the limit of $c_-=0$ are
\begin{align}
 h_T^+(c_+,0)& =\frac{\alpha}{\pi}\frac{1}{ec_+}\biggl\{-9\zeta'\bigl(-2,\tfrac{1}{2}\tfrac{m^2}{ec_+}\bigr) +6(\tfrac{1}{2}\tfrac{m^2}{ec_+})\zeta'\bigl(-1,\tfrac{1}{2}\tfrac{m^2}{ec_+}\bigr) \nonumber\\
 &\hspace*{2cm}+\frac{1}{4}\Bigl[1+2(\tfrac{1}{2}\tfrac{m^2}{ec_+})^2 \Bigr] (\tfrac{1}{2}\tfrac{m^2}{ec_+})
 -\Bigl[\frac{3}{2}(\tfrac{1}{2}\tfrac{m^2}{ec_+})-1\Bigr](\tfrac{1}{2}\tfrac{m^2}{ec_+})\ln(\tfrac{1}{2}\tfrac{m^2}{ec_+}) \biggr\}\,, \nonumber\\
 h_T^-(c_+,0)& =\frac{\alpha}{\pi}\frac{1}{ec_+}\biggl\{-\frac{27}{4}\zeta'\bigl(-2,\tfrac{1}{2}\tfrac{m^2}{ec_+}\bigr) +3(\tfrac{1}{2}\tfrac{m^2}{ec_+})\zeta'\bigl(-1,\tfrac{1}{2}\tfrac{m^2}{ec_+}\bigr) +\frac{3}{4}(\tfrac{1}{2}\tfrac{m^2}{ec_+})^2 \zeta'\bigl(0,\tfrac{1}{2}\tfrac{m^2}{ec_+}\bigr)  \nonumber\\
 &\hspace*{2cm}+\frac{1}{4}\Bigl[1+3(\tfrac{1}{2}\tfrac{m^2}{ec_+})^2 \Bigr] (\tfrac{1}{2}\tfrac{m^2}{ec_+})
 -\Bigl[\frac{3}{2}(\tfrac{1}{2}\tfrac{m^2}{ec_+})-\frac{5}{8}\Bigr](\tfrac{1}{2}\tfrac{m^2}{ec_+})\ln(\tfrac{1}{2}\tfrac{m^2}{ec_+})  \nonumber\\
 &\hspace*{2cm}
 +\frac{1}{4}(\tfrac{1}{2}\tfrac{m^2}{ec_+})\psi(\tfrac{1}{2}\tfrac{m^2}{ec_+})+\frac{1}{8} \biggr\}\,,\label{eq:explexpsh}\\
 h^+_{\bar F\bar F}(c_+,0)&=\frac{\alpha}{\pi}\frac{1}{ec_+} \frac{1}{c_+^2}\biggl\{3\zeta'\bigl(-2,\tfrac{1}{2}\tfrac{m^2}{ec_+}\bigr) +2\bigl(\tfrac{1}{2}\tfrac{m^2}{ec_+}\bigr)\zeta'\bigl(-1,\tfrac{1}{2}\tfrac{m^2}{ec_+}\bigr) -2(\tfrac{1}{2}\tfrac{m^2}{ec_+})^2 \zeta'\bigl(0,\tfrac{1}{2}\tfrac{m^2}{ec_+}\bigr)\nonumber\\
&\hspace*{2.5cm}  -\frac{1}{12}\Bigl[5+14(\tfrac{1}{2}\tfrac{m^2}{ec_+})^2\Bigr](\tfrac{1}{2}\tfrac{m^2}{ec_+}) +\Bigl[\frac{3}{2}(\tfrac{1}{2}\tfrac{m^2}{ec_+})-1\Bigr](\tfrac{1}{2}\tfrac{m^2}{ec_+})\ln(\tfrac{1}{2}\tfrac{m^2}{ec_+}) \nonumber\\ 
 &\hspace*{2.5cm} +\frac{1}{3}(\tfrac{1}{2}\tfrac{m^2}{ec_+})\psi(\tfrac{1}{2}\tfrac{m^2}{ec_+}) +\frac{1}{6} (\tfrac{1}{2}\tfrac{m^2}{ec_+})^2 \zeta\bigl(2,\tfrac{1}{2}\tfrac{m^2}{ec_+}\bigr)+\frac{1}{12} \biggr\}\,,\nonumber\\
  h^-_{\bar F\bar F}(c_+,0)& =\frac{\alpha}{\pi}\frac{1}{ec_+} \frac{1}{c_+^2} \biggl\{\frac{15}{4}\zeta'\bigl(-2,\tfrac{1}{2}\tfrac{m^2}{ec_+}\bigr) -\bigl(\tfrac{1}{2}\tfrac{m^2}{ec_+}\bigr)\zeta'\bigl(-1,\tfrac{1}{2}\tfrac{m^2}{ec_+}\bigr)+\frac{1}{4}(\tfrac{1}{2}\tfrac{m^2}{ec_+})^2 \zeta'\bigl(0,\tfrac{1}{2}\tfrac{m^2}{ec_+}\bigr) \nonumber\\
  &\hspace*{2.5cm}  -\frac{1}{6}\Bigl[3+3(\tfrac{1}{2}\tfrac{m^2}{ec_+})-\frac{5}{2}(\tfrac{1}{2}\tfrac{m^2}{ec_+})^2 \Bigr](\tfrac{1}{2}\tfrac{m^2}{ec_+})  +\Bigl[\frac{3}{2}(\tfrac{1}{2}\tfrac{m^2}{ec_+})-\frac{5}{8}\Bigr](\tfrac{1}{2}\tfrac{m^2}{ec_+})\ln(\tfrac{1}{2}\tfrac{m^2}{ec_+}) \nonumber\\
 &\hspace*{2.5cm}
 +\Bigl[\frac{1}{12}-(\tfrac{1}{2}\tfrac{m^2}{ec_+})^2\Bigr](\tfrac{1}{2}\tfrac{m^2}{ec_+})\psi(\tfrac{1}{2}\tfrac{m^2}{ec_+})
  +\frac{1}{6} (\tfrac{1}{2}\tfrac{m^2}{ec_+})^2 \zeta\bigl(2,\tfrac{1}{2}\tfrac{m^2}{ec_+}\bigr) -\frac{1}{24}\biggr\}\,, \nonumber
\end{align}
where $\psi(\cdot)$ denotes the digamma function.
Making use of the all-orders asymptotic expansions of the Hurwitz zeta function and its derivatives for large arguments, given, e.g., in Ref.~\cite{NIST}, it can be straightforwardly checked that \Eqref{eq:h+-B} is recovered from \Eqref{eq:explexpsh}.

The strong-field expansions of \Eqref{eq:explexpsh} follow from the series representations of the Hurwitz zeta function and its derivatives, cf., e.g., Refs.~\cite{Dunne:2004nc,Dowker:2015vya,Wolfram}. They read
\begin{align}
 h_T^+(c_+,0)& =\frac{\alpha}{\pi}\frac{1}{ec_+}\biggl\{ -9\zeta'(-2)+\Bigl[\ln\bigl(\tfrac{1}{2}\tfrac{m^2}{ec_+}\bigr)-12\zeta'(-1)-\frac{1}{2}\Bigr] \bigl(\tfrac{1}{2}\tfrac{m^2}{ec_+}\bigr)\nonumber\\
 &\hspace*{0.8cm}+\frac{3}{2}\Bigl[\ln\bigl(\tfrac{1}{2}\tfrac{m^2}{ec_+}\bigr)+\ln(2\pi)-\frac{5}{2}\Bigr] \bigl(\tfrac{1}{2}\tfrac{m^2}{ec_+}\bigr)^2-\bigl(\tfrac{1}{2}\tfrac{m^2}{ec_+}\bigr)^3 \nonumber\\
 &\hspace*{0.8cm}+6\sum_{j=0}^\infty (-1)^j\frac{j+1}{(j+2)(j+3)(j+4)} \zeta(j+2)\bigl(\tfrac{1}{2}\tfrac{m^2}{ec_+}\bigr)^{j+4}\biggr\}\,, \nonumber\\
 h_T^-(c_+,0)& =\frac{\alpha}{\pi}\frac{1}{ec_+}\biggl\{ -\frac{27}{4}\zeta'(-2)-\frac{1}{8}
 +\frac{1}{2}\Bigl[\frac{5}{4}\ln\bigl(\tfrac{1}{2}\tfrac{m^2}{ec_+}\bigr)-\frac{\gamma}{2}-21\zeta'(-1)-\frac{5}{8}\Bigr]\bigl(\tfrac{1}{2}\tfrac{m^2}{ec_+}\bigr)
  \nonumber\\
 &\hspace*{0.8cm}+\frac{3}{2}\Bigl[\ln\bigl(\tfrac{1}{2}\tfrac{m^2}{ec_+}\bigr) +\ln(2\pi) +\frac{\pi^2}{36}-\frac{19}{8}\Bigr]\bigl(\tfrac{1}{2}\tfrac{m^2}{ec_+}\bigr)^2 -\frac{1}{4}\Bigl[\frac{9}{2}+\zeta(3)\Bigr] \bigl(\tfrac{1}{2}\tfrac{m^2}{ec_+}\bigr)^3 \nonumber\\
 &\hspace*{0.8cm}
 +\frac{1}{4}\sum_{j=0}^\infty(-1)^j\Bigl[\frac{3(j^2+11j+10)}{(j+2)(j+3)(j+4)}\zeta(j+2)+\zeta(j+4)\Bigr]\bigl(\tfrac{1}{2}\tfrac{m^2}{ec_+}\bigr)^{j+4} \biggr\}\,,\label{eq:sfe_h}\\
 h^+_{\bar F\bar F}(c_+,0)&=\frac{\alpha}{\pi}\frac{1}{ec_+} \frac{1}{c_+^2}\biggl\{3\zeta'(-2)-\frac{1}{12}-\Bigl[\ln\bigl(\tfrac{1}{2}\tfrac{m^2}{ec_+}\bigr)+\frac{\gamma}{3}-8\zeta'(-1)+\frac{1}{6}\Bigr]\bigl(\tfrac{1}{2}\tfrac{m^2}{ec_+}\bigr)\nonumber\\
&\hspace*{0.8cm}  -\frac{1}{2}\Bigl[3\ln\bigl(\tfrac{1}{2}\tfrac{m^2}{ec_+}\bigr)+3\ln(2\pi)-\frac{\pi^2}{6}-\frac{13}{2}\Bigr]\bigl(\tfrac{1}{2}\tfrac{m^2}{ec_+}\bigr)^2 +\frac{2}{3}\Bigl[2-\zeta(3)\Bigr]\bigl(\tfrac{1}{2}\tfrac{m^2}{ec_+}\bigr)^3\nonumber\\ 
 &\hspace*{0.8cm} -\sum_{j=0}^\infty(-1)^j\Bigl[\frac{2(j^2+6j+5)}{(j+2)(j+3)(j+4)}\zeta(j+2)-\frac{j+5}{6}\zeta(j+4)\Bigr]\bigl(\tfrac{1}{2}\tfrac{m^2}{ec_+}\bigr)^{j+4} \biggr\}\,,\nonumber\\
  h^-_{\bar F\bar F}(c_+,0)& =\frac{\alpha}{\pi}\frac{1}{ec_+} \frac{1}{c_+^2} \biggl\{\frac{15}{4}\zeta'(-2)+\frac{1}{24}-\Bigl[\frac{5}{8}\ln \bigl(\tfrac{1}{2}\tfrac{m^2}{ec_+}\bigr)+\frac{\gamma}{12}-\frac{13}{2}\zeta'(-1)+\frac{3}{16}\Bigr] \bigl(\tfrac{1}{2}\tfrac{m^2}{ec_+}\bigr)\nonumber\\
  &\hspace*{0.8cm} -\frac{1}{2}\Bigl[3\ln \bigl(\tfrac{1}{2}\tfrac{m^2}{ec_+}\bigr)+3\ln(2\pi)-\frac{\pi^2}{12}-\frac{45}{8}\Bigr] \bigl(\tfrac{1}{2}\tfrac{m^2}{ec_+}\bigr)^2+\frac{1}{12}\Bigl[\frac{43}{2}-5\zeta(3)\Bigr] \bigl(\tfrac{1}{2}\tfrac{m^2}{ec_+}\bigr)^3\nonumber\\
 &\hspace*{0.8cm}
 -\frac{1}{4}\sum_{j=0}^\infty(-1)^j\Bigl[\frac{4j^3+35j^2+101j+70}{(j+2)(j+3)(j+4)}\zeta(j+2)-\frac{2j+7}{3}\zeta(j+4)\Bigr] \bigl(\tfrac{1}{2}\tfrac{m^2}{ec_+}\bigr)^{j+4} \biggr\}\,, \nonumber
\end{align}
where $\gamma$ is the Euler-Mascheroni constant, $\zeta(\cdot)$ is the Riemann zeta function, and $\zeta'(\cdot)$ is its derivative.
The analogous results determining \Eqref{eq:Gamma2derivs_B(E)} for a finite value of $c_-$ but $c_+=0$ follow from Eqs.~\eqref{eq:h+-B}, \eqref{eq:explexpsh} and \eqref{eq:sfe_h}
via the identity $h_p^\pm(0,c_-)=\pm h_p^\pm(c_+,0)|_{c_+\to-{\rm i}c_-}$ with $p\in\{T,FF\}$.

In the special case of a purely magnetic field $\vec{B}$, we have $c_+=|\vec{B}|=B$ and \Eqref{eq:Gamma2derivs_B(E)} can be expressed as
\begin{align}
 {\cal L}_{\rm HE}(\vec{B})\big|_{\sim \partial^2}&= -\frac{1}{2}\Bigl\{(\partial_0\vec{B})^2h_T^+(B,0) +[\vec{B}\cdot(\partial_0\vec{B})]^2h_{\bar F\bar F}^+(B,0) \nonumber\\
 &\hspace*{1.4cm}+(\partial_i\vec{B})^2\bigl[h_T^-(B,0)-h_T^+(B,0)\bigr] +[\vec{B}\cdot(\partial_i\vec{B})]^2\bigl[h_{\bar F\bar F}^-(B,0)-h_{\bar F\bar F}^+(B,0)\bigr] \nonumber\\
 &\hspace*{1.4cm}-\bigl[(\hat{\vec{B}}\cdot\vec{\nabla})\vec{B}\bigr]^2h_T^-(B,0)
-\bigl\{\vec{B}\cdot[(\hat{\vec{B}}\cdot\vec{\nabla})\vec{B}]\bigr\}^2h_{\bar F\bar F}^-(B,0)\Bigr\}\,,
\label{eq:Gamma2derivs_explB}
\end{align}
with $\vec{B}=B\hat{\vec{B}}$ and $|\hat{\vec{B}}|=1$. In \Eqref{eq:Gamma2derivs_explB} the Einstein summation convention over the index $i\in\{1,2,3\}$ is implicitly assumed.

On the other hand, it is well-known that the Heisenberg-Euler Lagrangian develops a manifestly non-perturbative imaginary part in electromagnetic fields for which $c_-\neq0$. The latter can be readily evaluated with the residue theorem. As obvious from Eqs.~\eqref{eq:pure_z_parallel} and \eqref{eq:pure_z_perp}, particularly for the case of $c_+=0$ this evaluation boils down to the use of the single identity
\begin{equation}
{\rm Im}\Bigl\{{\rm i}\frac{\alpha}{2\pi}\int_0^\infty{\rm d}s\,{\rm e}^{-{\rm i}m^2s}\,g(z)\cot z\big|_{z\to-{\rm i}z'}\Bigl\}\,=\frac{\alpha}{2}\frac{1}{ec_-}\sum_{n=1}^\infty {\rm e}^{-\frac{m^2}{ec_-}n\pi}\,g(-n\pi)\,, \label{eq:ImId}
\end{equation}
where $g(z)$ is an analytic function: all expressions $g(z)\cot z$ to be considered here are regular at $z\to0$ such that there is no pole at $z=0\,\leftrightarrow\,n=0$; cf. \Eqref{eq:calN_series}.
Therewith, we infer
\begin{align}
{\rm Im}\bigl\{h_T^+(0,c_-)\bigr\}&= \frac{\alpha}{4}\frac{1}{ec_-}\sum_{n=1}^\infty {\rm e}^{-\frac{m^2}{ec_-}n\pi}\biggl[\frac{m^2}{ec_-} +\frac{3}{n\pi}\biggr]\frac{3}{(n\pi)^2}\,, \nonumber\\
{\rm Im}\bigl\{h_T^-(0,c_-)\bigr\}&= \frac{\alpha}{4}\frac{1}{ec_-}\sum_{n=1}^\infty {\rm e}^{-\frac{m^2}{ec_-}n\pi}\biggl[\Bigl(\frac{m^2}{ec_-}\Bigr)^2\frac{3}{8n\pi}+\frac{1}{2}\frac{m^2}{ec_-}\Bigl(1-\frac{3}{(n\pi)^2}\Bigr)-\frac{27}{4(n\pi)^3}\biggr]\,,\label{eq:Imhc+} \\
{\rm Im}\bigl\{h_{\bar F\bar F}^+(0,c_-)\bigr\}&= -\frac{\alpha}{4}\frac{1}{ec_-} \frac{1}{c_-^2}\sum_{n=1}^\infty {\rm e}^{-\frac{m^2}{ec_-}n\pi}\biggl[\Bigl(\frac{m^2}{ec_-}\Bigr)^2\Bigl(\frac{n\pi}{3}+\frac{1}{n\pi}\Bigr)-\frac{m^2}{ec_-}\Bigl(\frac{2}{3}-\frac{1}{(n\pi)^2}\Bigr)-\frac{3}{(n\pi)^3}\biggr]\,, \nonumber\\
{\rm Im}\bigl\{h_{\bar F\bar F}^-(0,c_-)\bigr\}&=-\frac{\alpha}{4}\frac{1}{ec_-} \frac{1}{c_-^2} \sum_{n=1}^\infty {\rm e}^{-\frac{m^2}{ec_-}n\pi}\biggl[\frac{1}{2}\Bigl(\frac{m^2}{ec_-}\Bigr)^3-\Bigl(\frac{m^2}{ec_-}\Bigr)^2\Bigl(\frac{n\pi}{3}-\frac{1}{8n\pi}\Bigr)\nonumber\\
&\hspace*{5cm}+ \frac{1}{2}\frac{m^2}{ec_-}\Bigl(\frac{1}{3}+\frac{1}{(n\pi)^2}\Bigr)+\frac{15}{4(n\pi)^3}\biggr]\,. \nonumber
\end{align}
These expressions constitute the imaginary part of \Eqref{eq:Gamma2derivs_B(E)} for $c_+=0$ and result in corrections to the Schwinger-formula describing the decay of the quantum vacuum via electron-positron pair production in slowly-varying electric fields: the leading derivative correction to the vacuum decay rate $w(\bar F)=2\,{\rm Im}\{{\cal L}_{\rm HE}(\bar F)\}$ \cite{Heisenberg:1935qt,Schwinger:1951nm} is given by $w(\bar F)|_{\sim\partial^2}=2\,{\rm Im}\{{\cal L}_{\rm HE}(\bar F)|_{\sim\partial^2}\}$; cf. also Refs.~\cite{Dittrich:1985yb,Dunne:2004nc, Cohen:2008wz} and references therein.

Especially for a purely electric field $\vec{E}$ we have $c_-=E$, such that \Eqref{eq:Gamma2derivs_B(E)} becomes
\begin{align}
 {\cal L}_{\rm HE}(\vec{E})\big|_{\sim \partial^2}&= \frac{1}{2}\Bigl\{(\partial_0\vec{E})^2\bigl[h_T^-(0,E)+h_T^+(0,E)\bigr] -[\vec{E}\cdot(\partial_0\vec{E})]^2\bigl[h_{\bar F\bar F}^-(0,E)+h_{\bar F\bar F}^+(0,E)\bigr]\nonumber\\
 &\hspace*{1.4cm} -(\partial_i\vec{E})^2h_T^+(0,E) +[\vec{E}\cdot(\partial_i\vec{E})]^2h_{\bar F\bar F}^+(0,E)\nonumber\\
 &\hspace*{1.4cm}-\bigl[(\hat{\vec{E}}\cdot\vec{\nabla})\vec{E}\bigr]^2h_T^-(0,E)
+\bigl\{\vec{E}\cdot[(\hat{\vec{E}}\cdot\vec{\nabla})\vec{E}]\bigr\}^2h_{\bar F\bar F}^-(0,E)\Bigr\}\,.
\label{eq:Gamma2derivs_explE}
\end{align}

The associated derivative correction to the vacuum decay rate is $w(\vec{E})|_{\sim\partial^2}=2\,{\rm Im}\{{\cal L}_{\rm HE}(\vec{E})|_{\sim\partial^2}\}$.
A comparison of \Eqref{eq:Gamma2derivs_explE} with \Eqref{eq:Gamma2derivs_explB} implies that
\begin{equation}
 {\cal L}_{\rm HE}(\vec{E})\big|_{\sim \partial^2}= -{\cal L}_{\rm HE}(\vec{B})\big|_{\sim \partial^2}\Big|_{B\to-{\rm i}E,\partial_0\leftrightarrow\partial_i}\,.
 \label{eq:Gamma2derivs_explE2}
\end{equation}
Recall that $h_p^\pm(B,0)\big|_{B\to-{\rm i}E}= \pm h_p^\pm(0,E) $.

It can be straightforwardly checked that for the special cases considered explicitly by Refs.~\cite{Lee:1989vh,Gusynin:1998bt}, namely either a purely magnetic field directed along the $z$ axis which only depends on $x$ and $y$, or a purely electric field directed along the $x$ axis which exclusively depends on $t$ and $x$, the known results are recovered.
In fact, the non-trivial structures of the effective Lagrangians associated with these cases are fully determined by 
$h_T^-(B,0)- h_T^+(B,0)+B^2[h_{\bar F\bar F}^-(B,0)- h_ {\bar F\bar F}^+(B,0)]\sim\int_0^\infty{\rm d}s\,{\rm e}^{-{\rm i}m^2s}\frac{z}{\sin z}\,{\cal N}_2^\perp(z,0)$ for the magnetic field $\vec{B}=B(x,y)\,\vec{e}_{\rm z}$, and similarly
$h_T^-(0,E)+ h_T^+(0,E)-E^2[h_{\bar F\bar F}^-(0,E)+ h_ {\bar F\bar F}^+(0,E)]\sim\int_0^\infty{\rm d}s\,{\rm e}^{-{\rm i}m^2s}\frac{z'}{\sinh z'}\,{\cal N}_1^\parallel(0,z')$ for the electric field $\vec{E}=E(t,x)\,\vec{e}_{\rm x}$.

Finally, we note that in the limit of crossed fields of the same amplitude characterized by $\vec{E}(x)\cdot\vec{B}(x)=0$ and $|\vec{E}(x)|=|\vec{B}(x)|$, we have $c_+=c_-={\cal F}={\cal G}=0$. Because of $\partial_\alpha\bar F_{\sigma\mu}\bar F^{\sigma\mu}=2\partial_\alpha {\cal F}=0$, in this case
\Eqref{eq:Gamma2derivs_B(E)} takes an especially simple form, namely
\begin{align}
 {\cal L}_{\rm HE}(\bar F)\big|_{\sim \partial^2}=-\frac{1}{4}\partial_\alpha\bar F_{\mu\nu} \partial_\beta\bar F^{\mu\nu}\, h^{\alpha\beta}_T(\bar F) \,.
 \label{eq:Gamma2derivs_c+c-=0_v0}
\end{align}
Accounting for the fact that in this limit the tensor structure $h^{\alpha\beta}_T(\bar F)$ can be compactly represented as, cf. Eqs.~\eqref{eq:h_p}, \eqref{eq:weakfield}
and \eqref{eq:h+-B},
\begin{equation}
 h^{\alpha\beta}_T(\bar F) = \frac{1}{15} \frac{\alpha}{\pi} \frac{1}{m^2}\Bigl(\bar F^\alpha_{\ \, \tau}\bar F^{\beta\tau}\frac{1}{7}\Bigl(\frac{e}{m^2}\Bigr)^2 -g^{\alpha\beta}\Bigr)\,,
\end{equation}
\Eqref{eq:Gamma2derivs_c+c-=0_v0} becomes
\begin{align}
 {\cal L}_{\rm HE}(\bar F)\big|_{\sim \partial^2}=\frac{1}{60} \frac{\alpha}{\pi} \frac{1}{m^2}\Bigl[\partial_\alpha\bar F_{\mu\nu} \partial^\alpha\bar F^{\mu\nu}-\frac{1}{7}\Bigl(\frac{e}{m^2}\Bigr)^2\partial_\alpha\bar F_{\mu\nu} \partial_\beta\bar F^{\mu\nu}\bar F^\alpha_{\ \, \tau}\bar F^{\beta\tau}\Bigr]\,.
 \label{eq:Gamma2derivs_c+c-=0}
\end{align}
As to be expected, this expression vanishes identically in plane wave fields \cite{Schwinger:1951nm}.

\section{Conclusions and Outlook}\label{sec:concls}

In this work, we put forward an alternative way to evaluate derivative corrections to the Heisenberg-Euler effective action in slowly varying electromagnetic fields. Using the explicit results available in the literature for the one-loop vacuum polarization tensor in the presence of a constant electromagnetic field as central input, we arrive at a rather compact expression for the quadratic derivative correction to the Heisenberg-Euler effective action at one loop.

For the special cases of magnetic- and electric-like field configurations characterized by the vanishing of one of the secular invariants of the electromagnetic field, we obtain closed-form expressions and work out all-orders weak- and strong-field expansions.

Apart from providing insights into fundamental aspects of strong-field QED, our results are relevant for precision studies of quantum vacuum nonlinearities in experimentally realistic field configurations beyond the locally constant field approximation.

Of course, the strategy devised in the present work to determine derivative corrections to the Heisenberg-Euler effective action for QED in four space-time dimensions can be readily extended to QED in other space-time dimensions as well as to other field theories, such as scalar QED; cf also Ref.~\cite{Gusynin:1998bt}.

\acknowledgments

This work has been funded by the Deutsche Forschungsgemeinschaft (DFG) under Grant No. 416607684 within the Research Unit FOR2783/1. In memoriam \texttt{Maria Rohrmeier (12.8.1981 - 2.8.2021)}.

\appendix

\section{Identities}

Starting from the identity ${}^*\!\bar F^{\mu\alpha}\bar F^\nu_{\ \alpha}={\cal G}g^{\mu\nu}$ it can be easily shown that ${\cal G}^2$ can be expressed in terms of the field strength tensor alone, without resorting to the dual field strength tensor or expressions involving the Levi-Civita symbol, respectively.
More specifically, we have
\begin{align}
 {\cal G}^2=\frac{1}{4}({}^*\!\bar F_{\mu\alpha}\bar F^{\nu \alpha})(\bar F_{\nu\beta}{}^*\!\bar F^{\mu\beta})=-\frac{1}{8}(\bar F_{\rho\sigma}\bar F^{\rho\sigma}\bar F_{\beta\nu}\bar F^{\beta\nu}-2\bar F_{\rho\sigma}\bar F^{\rho\nu}\bar F_{\beta\nu}\bar F^{\beta\sigma})\,.
 \label{eq:id:G^2}
\end{align}
The last identity follows straightforwardly upon plugging in the definition of the dual field strength tensor and making use of the fact that
\begin{equation}
 \epsilon^{\mu\alpha\rho\sigma}\epsilon_{\mu\beta\kappa\lambda}=-(\delta_\beta^\alpha\delta_\kappa^\rho\delta_\lambda^\sigma+\delta_\kappa^\alpha\delta_\lambda^\rho\delta_\beta^\sigma+\delta_\lambda^\alpha\delta_\beta^\rho\delta_\kappa^\sigma-\delta_\kappa^\alpha\delta_\beta^\rho\delta_\lambda^\sigma-\delta_\beta^\alpha\delta_\lambda^\rho\delta_\kappa^\sigma-\delta_\lambda^\alpha\delta_\kappa^\rho\delta_\kappa^\beta)\,.
\end{equation}

Along the same lines, we can express the scalar quantity $(\partial_\alpha{\cal G})(\partial^\alpha{\cal G})$ as
\begin{align}
 \partial_\rho{\cal G}\partial^\rho{\cal G}&=\frac{1}{4}\partial_\rho({}^*\!\bar F_{\mu\alpha}\bar F^{\nu \alpha})\partial^\rho(\bar F_{\nu\beta}{}^*\!\bar F^{\mu\beta}) \nonumber\\
 &=\bar F_{\rho\mu}\bar F^{\sigma\mu}\partial_\alpha\bar F_{\sigma\nu}\partial^\alpha\bar F^{\rho\nu}
 -\frac{1}{4}\bar F_{\mu\nu}\bar F^{\mu\nu}\partial_\alpha\bar F_{\rho\sigma} \partial^\alpha\bar F^{\rho\sigma}
 -\frac{1}{4}\bar F_{\mu\nu}\partial_\alpha\bar F^{\mu\nu}\bar F_{\rho\sigma}\partial^\alpha\bar F^{\rho\sigma}\,.
\end{align}
Taking into account the obvious fact that
\begin{equation}
 \partial_\rho{\cal G}\partial^\rho{\cal G}=\frac{1}{4}{}^*\!\bar F_{\mu\nu}\partial_\rho\bar F^{\mu\nu} {}^*\!\bar F_{\alpha\beta}\partial^\rho\bar F^{\alpha\beta}\,,
\end{equation}
we can infer the following identity
\begin{align}
 {}^*\!\bar F_{\mu\nu}\partial_\rho\bar F^{\mu\nu} {}^*\!\bar F_{\alpha\beta}\partial^\rho\bar F^{\alpha\beta}
 &=4\bar F_{\rho\mu}\bar F^{\sigma\mu}\partial_\alpha\bar F_{\sigma\nu}\partial^\alpha\bar F^{\rho\nu}
 -\bar F_{\mu\nu}\bar F^{\mu\nu}\partial_\alpha\bar F_{\rho\sigma} \partial^\alpha\bar F^{\rho\sigma} \nonumber\\
 &\quad-\bar F_{\mu\nu}\partial_\alpha\bar F^{\mu\nu}\bar F_{\rho\sigma}\partial^\alpha\bar F^{\rho\sigma}\,.
 \label{eq:id:*F*FFF}
\end{align}

\end{document}